\documentclass[aip,graphicx]{revtex4-1}

\draft 

\usepackage[pdfencoding=auto]{hyperref}

\usepackage{color}
\usepackage{amsmath,amssymb}
\usepackage{mathrsfs}  
\usepackage{braket}
\usepackage{bbm}
\usepackage{enumitem}

\usepackage{mathtools}
\usepackage[makeroom]{cancel}
\usepackage{tensor}

\usepackage{amsthm}
\theoremstyle{plain}
\newtheorem{thm}{Theorem}
\newtheorem{prop}{Proposition}
\newtheorem{lemm}{Lemma}

\theoremstyle{definition}

\newcommand{\defarrow}{\stackrel{\mathrm{def.}}{\Leftrightarrow}}

\newcommand{\natn}{\mathbb{N}}

\newcommand{\realn}{\mathbb{R}}
\newcommand{\cmplx}{\mathbb{C}}

\newcommand{\unit}{\mathbbm{1}}
\DeclareMathOperator{\tr}{tr}

\newcommand{\cL}{\mathcal{L}}

\newcommand{\cI}{\mathcal{I}}
\newcommand{\cH}{\mathcal{H}}

\newcommand{\LH}{\cL(\cH)}
\newcommand{\LLH}{\cL(\cL(\cH))}

\newcommand{\LX}{\mathcal{L}(X)}

\newcommand{\cph}{\mathcal{CP}(\mathcal{H})}

\newcommand{\dt}{\delta t}

\newcommand{\ns}[1]{\tensor*[^\ast]{#1}{}}
\newcommand{\nsfin}[1]{\tensor*[^{\ast}_{}]{#1}{^{\mathrm{fin}}_{}}}
\DeclareMathOperator{\st}{st}

\newcommand{\tM}{\widetilde{M}}

\newcommand{\rmax}{\mathrm{max}}
\newcommand{\rmin}{\mathrm{min}}

\begin{document}


\title[Nonstandard derivation of the GKSL master equation from the Kraus representation]{Nonstandard derivation of the Gorini-Kossakowski-Sudarshan-Lindblad master equation of a quantum dynamical semigroup from the Kraus representation}
\author{Yui Kuramochi}
\email{yui.tasuke.kuramochi@gmail.com}
 \altaffiliation{Graduate School of Informatics and Engineering, The University of Electro-Communications, 1-5-1 Chofugaoka, Chofu, Tokyo 182-8585, Japan}

\date{\today}





\begin{abstract}
We give a new nonstandard proof of the well-known theorem that the generator $L$ of a quantum dynamical semigroup $\exp(tL)$ on a finite-dimensional quantum system has a specific form called a Gorini-Kossakowski-Sudarshan-Lindblad (GKSL) generator (also known as a Lindbladian) and vice versa.
The proof starts from the Kraus representation of the quantum channel $\exp (\delta t L)$ for an infinitesimal hyperreal number $\delta t>0$ and then estimates the orders of the traceless components of the Kraus operators.
The jump operators naturally arise as the standard parts of the traceless components of the Kraus operators divided by $\sqrt{\delta t}$.
We also give a nonstandard proof of a related fact that close completely positive maps have close Kraus operators.
\end{abstract}


\maketitle 




%
%

%


\section{Introduction} \label{sec:intro}

Let $\cH$ be a finite dimensional complex Hilbert space and let $\LH$ and $\LLH$ denote the sets of linear operators on $\cH$ and of linear maps on $\LH$, respectively.  
A parametrized family $(\Lambda_t)_{t\geq 0}$ in $\LLH$ is called a one-parameter semigroup on $\LH$ if it satisfies the following conditions:
\begin{enumerate}[label=(\roman*)]
\item
$\Lambda_0 = \cI$ (the identity map on $\LH$).
\item
$\Lambda_{t+s} = \Lambda_t \circ \Lambda_s$ $(t,s \geq 0)$.
\item
$[0,\infty) \ni t \mapsto \Lambda_t \in \LLH$ is continuous.
\end{enumerate}
According to the general theory (e.g.\ Ref.~\onlinecite{davies1980}, Lemma~1.1) any one-parameter semigroup $(\Lambda_t)_{t\geq 0}$ on $\LH$ has a generator $L \in \LLH$ such that $\Lambda_t = \exp (tL)$.
Here the domain of $L$ is the total space $\LH$ due to the finite-dimensionality of $\LH$.

A one-parameter semigroup $(\Lambda_t)_{t\geq 0}$ on $\LH$ is called a \textit{quantum dynamical semigroup (QDS)} \cite{Gorini1976,lindblad1976,davies1976quantum} if it further satisfies 
\begin{enumerate}[label=(\roman*),resume]
\item\label{it:4}
$\Lambda_t$ is completely positive (CP) \cite{1955stinespring,Kraus1971311,davies1976quantum,kraus10.1007/3-540-12732-1} for all $t\geq 0$.
\item \label{it:5}
$\Lambda_t (\unit) = \unit$ for all $t\geq 0$, where $\unit$ is the unit operator on $\cH$.
\end{enumerate}
The conditions~\ref{it:4} and \ref{it:5} mean that $\Lambda_t$ is a quantum channel in the Heisenberg picture.

The celebrated characterization of the generator of a quantum dynamical semigroup is stated as follows:
\begin{thm}[Refs.~\onlinecite{Gorini1976,lindblad1976}] \label{thm:gksl}
The generator $L$ of a QDS $(\Lambda_t)_{t\geq 0}$ on $\LH$ has the following form:
\begin{equation}
	L(A) = i[H,A] + \sum_{j=1}^{N} \left(V_j^\dag A V_j - \frac{1}{2} \{V_j^\dag V_j ,A\} \right) \quad (A \in \LH), \label{eq:gksl1}
\end{equation}
where $N \geq 1$ is an integer, $\dag$ denotes the adjoint,  $H  \in \LH$ is self-adjoint, $V_j \in \LH$ $(j=1,\dots , N)$, $[A,B]:= AB - BA$, and $\{A,B\}:=AB+BA$.
The operators $H$ and $V_j$ can be chosen to be traceless and $N$ can be chosen to be at most $d^2-1$, where $d = \dim \cH$.

Conversely if $L \in \LLH$ is given by \eqref{eq:gksl1} for some $V_j \in \LH$ $(j=1,\dots, N)$ and some self-adjoint $H \in \LH$, then $(\exp (tL))_{t\geq 0}$ is a QDS.
\end{thm}
A generator given by \eqref{eq:gksl1} is called a \textit{Gorini-Kossakowski-Sudarshan-Lindblad (GKSL) generator} (or a \textit{Lindbladian}) and the operators $V_j$\rq{}s are called \textit{jump operators}.
For a given initial density operator $\rho \in \LH$, the time evolution of the system in the Schr\"odinger picture is given by $\rho(t) = \Lambda_t^\dag (\rho)$, where the adjoint map $\Psi^\dag \in \LLH$ of $\Psi \in \LLH$ is defined by $\tr[\Psi(A)\rho] = \tr[A \Psi^\dag(\rho)]$ $(A,\rho \in \LH)$ and here $\tr[\cdot]$ is the trace.
$\rho(t)$ satisfies the following \textit{GKSL master equation}
\begin{align*}
	\frac{d}{dt}\rho(t) & = L^\dag (\rho(t)) \\
	&= -i[H,\rho(t)]+  \sum_{j=1}^{N} \left(V_j \rho(t) V_j^\dag - \frac{1}{2} \{V_j^\dag V_j ,\rho(t) \} \right). \end{align*}
For a historical account of this equation, see Ref.~\onlinecite{chruscinski2017}.

The nontrivial part of the proof of Theorem~\ref{thm:gksl} is to show that any QDS has a GKSL generator, especially how to take into account the assumption of the complete positivity.
In the original proofs, this is done by considering a operator-valued positive-semidefinite bilinear form called the dissipator~\cite{lindblad1976} or using a characterization~\cite{kossakowski1972necessary} of a positive dynamical semigroup~\cite{Gorini1976}.

There is another heuristic short derivation of the GKSL generator starting from the well-known Kraus representation \cite{Kraus1971311,davies1976quantum,kraus10.1007/3-540-12732-1} of a CP map, which can be found for example in Ref.~\onlinecite{preskill1998lecture}.
Let us briefly see the derivation.
We consider a Kraus representation of $\Lambda_{dt}$ for an infinitesimal time $dt > 0$ as 
\begin{equation}
	\Lambda_{dt}(A) = \sum_{j=0}^{d^2-1} M_j(dt)^\dag A M_j(dt)
	\notag 
\end{equation}
and \textit{assume} that $M_j (dt)$ has the following form:
\begin{equation}
	M_j(dt) = \begin{cases}
	\unit + (G-iH)dt + o(dt) & (j=0); \\
	\sqrt{dt} V_j + O(dt) & (j=1,\dots,d^2-1), 
	\end{cases}
	\label{eq:assumption}
\end{equation}
where $G$ and $H$ are self-adjoint.
Taking the $O(dt)$ terms of the completeness condition $\Lambda_{dt}(\unit) = \sum_{j=0}^{d^2-1} M_j(dt)^\dag M_j(dt) = \unit$, we obtain
\begin{equation}
	G = -\frac{1}{2} \sum_{j=1}^{d^2-1} V_j^\dag V_j.
	\label{eq:G}
\end{equation}
From \eqref{eq:assumption} and \eqref{eq:G}, direct calculations show that $L = \lim_{dt\to +0} (\Lambda_{dt} - \mathcal{I})/dt$ is a GKSL generator \eqref{eq:gksl1}.

If we try to interpret the above argument as a proof of the direct part of Theorem~\ref{thm:gksl}, we need to justify the assumption \eqref{eq:assumption}.
For example we have to explain why the Kraus operators does not contain $O(\sqrt{dt}(\log dt)^{1976})$ terms.
The assumption also implies that the Kraus operators $M_j(dt)$ converges to $\delta_{j0}\unit$ when $dt \to +0$, while such continuity of Kraus operators is not obvious because the existence theorem of the Kraus representation say nothing about the relation between the Kraus operators of $\Lambda_t$ for different $t$\rq{}s.
Due to such difficulty or not, to the author\rq{}s knowledge, no rigorous proof of Theorem~\ref{thm:gksl} in this line is known.

In this paper, we give a new proof of Theorem~\ref{thm:gksl} which is based on the \textit{nonstandard analysis}~\cite{davis1977,loebwolff2015} and starts from the the Kraus representation of $\Lambda_{\delta t}$ for an infinitesimal hyperreal number $\delta t >0$.
The crucial step of the proof is the order estimation of the traceless components of Kraus operators, which finally justifies the above assumption~\eqref{eq:assumption} with some modifications of the definitions of operators (see \eqref{eq:4}, \eqref{eq:NjO} and \eqref{eq:ZO} in the proof).
Then the jump operators are naturally introduced as the standard parts of the traceless components of the Kraus operators divided by $\sqrt{\dt}$ (see \eqref{eq:jdef}).
We also prove that close CP maps have close Kraus operators, which justifies the above-mentioned nontrivial consequence of the assumption in a more general setting.

The resf of this paper is organized as follows.
In Section~\ref{sec:prel} we give some preliminaries on the nonstandard analysis and CP maps used in the proof.
In Section~\ref{sec:proof} we prove Theorem~\ref{thm:gksl}.
In Section~\ref{sec:rem} we make some remarks on the proof.
In Appendix~\ref{sec:app}, we give a nonstandard proof of Proposition~\ref{prop:conti} which claims that Kraus operators of close CP maps can be chosen to be close.

\section{Preliminaries} \label{sec:prel}
\subsection{Notation}
We denote by $\natn := \{1,2,\dots\}$, $\realn$, and $\cmplx$ the sets of natural, real, and complex numbers, respectively.
In this paper we denote by $\braket{\psi|\phi}$ the inner product on $\cH$ for $\psi, \phi \in \cH$, which is antilinear and linear with respect to $\psi$ and $\phi$, respectively.
For $\psi, \phi \in \cH$, the operator $\ket{\psi}\bra{\phi} \in \LH$, called the von Neumann-Schatten product, is defined by
\[
	(\ket{\psi}\bra{\phi} ) \xi := \braket{\phi|\xi} \psi \quad (\xi \in \cH).
\]
In this convention the appearances of equations become similar to the ones in  Dirac\rq{}s braket notation in physics.
(E.g.\ $\braket{\psi|A|\phi}$ in the braket notation is $\braket{\psi|A\phi}$ in our notation, which means the inner product of $\psi$ and $A\phi$.)

\subsection{Nonstandard analysis}
Here we summarize facts on nonstandard analysis used in the main proof.
For the full exposition of the theory, the reader is referred to any textbook on the nonstandard analysis (e.g.\ Refs.~\onlinecite{davis1977,loebwolff2015}).

Let $S$ be an infinite set that we want to investigate.
In this paper we assume $\natn, \realn, \cmplx, \cH \subseteq S$.
We consider a set $V(S)$ called a superstructure (or universe) defined via
\begin{gather*}
	V_0(S) := S, \quad V_{n+1}(S) := V_n(S) \cup \mathscr{P}(V_n(S)) \quad (n\geq 0), \\
	V (S):= \bigcup_{n\geq 0}V_n(S),
\end{gather*}
where $\mathscr{P}(X)$ denotes the power set of $X$.
The superstructure $V(S)$ contains $S$, $\mathscr{P}(S)$, any map $f\colon A \to B$ with $A, B \in V$, any set of maps from $A$ to $B$ with $A,B \in V$, $\LH$, $\LLH$, and so on.
(Note that a map in the set theory is identified with its graph which is also a set.)
In this sense $V(S)$ is sufficiently large so that all the objects we are interested in the ordinary mathematical study about $S$ are in $V(S)$.

In the nonstandard analysis, we take another superstructure $V(\ns{S})$ and a map $V(S) \ni x \mapsto \ns{x} \in V(\ns{S})$, called a $\ast$-map, satisfying the \textit{transfer principle}, which states that any sentence (closed logical formula) $\psi$ with constant terms in $V(S)$ is true if and only if the corresponding sentence $\ns{\psi}$, which is obtained by replacing each constant term $c$ appearing in $\psi$ with $\ns{c}$, is true.
From the transfer principle it follows that the $\ast$-map is injective and by identifying each $x \in V(S)$ with $\ns{x} \in V(\ns{S})$ we regard $V(S) \subseteq V(\ns{S})$.
An element in $V(S)$ is said to be \textit{standard}, while one in $V(\ns{S}) \setminus V(S)$ is said to be \textit{nonstandard}.
For constant real or complex number $a$ and a map $f \colon A \to B$ (e.g.\ we consider in this paper the matrix exponential $\exp(\cdot)$ on $\LH$) in $V(S)$, the stars in $\ns{a}$ and $\ns{f}$ are occasionally omitted as $a$ and $f$, while we do not abbreviate the stars of sets like $\ns{\natn}, \ns{\realn}, \ns{\cmplx}, \ns{\cH}, \ns{\LH},$ and $\ns{\LLH}$.
The spaces $\ns{\cH}$, $\ns{\LH}$, and $\ns{\LLH}$ are vector spaces over $\ns{\cmplx}$.
Elements of $\ns{\natn}, \ns{\realn}, \ns{\cmplx}$, and $\ns{\LH}$ are occasionally called  hypernatural, hyperreal, and hypercomplex numbers and $\ast$-operators, respectively.

Another important property to be required is that $V(\ns{S})$ is an \textit{enlargement} of $V(S)$ in the following sense.
A binary relation $R$ is said to be \textit{concurrent} if for any finite elements $a_1, \dots, a_n $ in the domain of $R$, there exists $b$ such that $\braket{a_j,b} \in R$ for all $j=1,\dots ,n$, where $\braket{x,y} $ denotes the ordered pair.
$V(\ns{S})$ is called an enlargement of $V(S)$ if for each concurrent binary relation $R \in V(S)$ there exists $\beta \in V(\ns{S})$ such that $\braket{\ns{a},\beta} \in \ns{R}$ for all $a$ in the domain of $R$. 
If $V(\ns{S})$ is an enlargement, then it follows that for any infinite set $A$ the $\ast$-set $\ns{A}$ contains a nonstandard element.
For given $S$ we can show that such a nonstandard model $(V(\ns{S}) , \ast)$ always exists and from now on we assume the above properties.

From the second requirement of the enlargement, we can also show the existence of the infinite and infinitesimal real numbers, which are defined as follows: For $\alpha \in \ns{\realn}$
\begin{itemize}
\item
$\alpha$ is \textit{finite} $:\defarrow$ there exists $a\in \realn$ such that $|\alpha | < a$,
\item
$\alpha$ is \textit{infinite} $:\defarrow$ $\alpha > a$ for all $a\in \realn$,
\item
$\alpha$ is \textit{infinitesimal} $:\defarrow$ $|\alpha|<a$ for all $a\in (0,\infty)$.
\end{itemize}
Note that rigorously speaking $<$ and $|\cdot|$ in the above definition are respectively $\ns{<}$ and $\ns{|\cdot|}$, while $\ast$\rq{}s are abbreviated for convenience.
A hypercomplex number $\alpha \in \ns{\cmplx}$ is said to be finite (respectively, infinitesimal) if $|\alpha|$ is finite (respectively, infinitesimal).
For $\alpha, \beta \in \ns{\cmplx}$, we write as $\alpha \approx \beta$ if $\alpha -\beta$ is infinitesimal.
We write the set of finite hyperreal and hypercomplex numbers as $\nsfin{\realn}$ and $\nsfin{\cmplx}$, respectively.
For each finite $ \alpha \in \nsfin{\realn}$ (respectively, $\alpha \in \nsfin{\cmplx}$) there exists a unique standard element $a \in \realn$ (respectively, $a\in \cmplx$) such that $\alpha \approx a$.
Such $a$ is called the \textit{standard part} of $\alpha$ and written as $\st (\alpha)$.
For $\alpha, \beta \in \nsfin{\cmplx}$ we have
\[
	\st(\alpha+\beta) = \st(\alpha) + \st(\beta), \quad
	\st(\alpha\beta) = \st(\alpha)  \st(\beta), \quad
	\st(\overline{\alpha}) = \overline{\st(\alpha)}
\]
and $\st(a) = a$ for $a\in \cmplx$.

The above notions are generalized to vectors.
Let $X \in V(S)$ be a finite-dimensional vector space over $\cmplx$ with a basis $(e_i)_{i=1}^n$.
In this paper, we consider the cases $X= \cH, \LH$, and $\LLH$.
By the transfer principle, each $\xi \in \ns{X}$ is uniquely written as a $\ast$-linear combination
\begin{equation}
	\xi = \sum_{j=1}^n \alpha_j e_j \quad (\alpha_1,\dots ,\alpha_n \in \ns{\cmplx}), 
	\label{eq:xi}
\end{equation}
and so $\ns{X}$ is isomorphic to $(\ns{\cmplx})^n$ as a vector space over $\ns{\cmplx}.$
We say $\xi \in \ns{X}$ is finite (respectively, infinitesimal) if $\|\xi\|$ is finite (respectively, infinitesimal), where $\|\cdot \|$ is a norm on $X$.
This definition is independent from the choice of the norm $\| \cdot \|$ because due to the finite-dimensionality of $X$ the norm $\| \cdot \|$ is equivalent to any other norm $\|\cdot\|^\prime$ on $X$ in the sense that there are real numbers $a,b >0$ satisfying
\[
	a \|x\| \leq \|x\|^\prime \leq b\|x\| \quad (\forall x\in X)
\]
(\onlinecite{conway1990}, Theorem~III.3.1).
If $\xi \in \ns{X}$ has the expression~\eqref{eq:xi}, then $\xi$ is finite (respectively, infinitesimal) if and only if each $\alpha_j$ is finite (respectively, infinitesimal).
For $\xi, \eta \in \ns{X}$, we write as $\xi \approx \eta$ if $\xi - \eta$ is infinitesimal.
For finite $\xi \in \ns{X}$, there exists a unique standard element $x\in X$, called the standard part of $\xi$, such that $\xi \approx x$. 
As in the scalar case such $x$ written as $\st(\xi)$.
If a finite element $\xi \in \ns{X}$ has the expression~\eqref{eq:xi}, then the standard part is given by
\[
	\st(\xi) = \sum_{j=1}^n \st(\alpha_j) e_j .
\]

Let $\alpha \in \ns{\realn}$ and let $\xi, \eta \in \ns{X}$.
We introduce the following order notation:
\begin{itemize}
\item
$\xi = O(\alpha)$ (or $O(\alpha)=\xi$) $:\defarrow$ there exists $0\leq a  \in \realn$ such that $\| \xi \| \leq a|\alpha|$.
\item
$\xi = \eta + O(\alpha)$ (or $\xi+O(\alpha)= \eta$) $:\defarrow$ $\xi -\eta = O(\alpha)$.
\end{itemize}
These definitions are again independent from the choice of the norm.
By this notation $\xi \in \ns{X}$ is finite if and only if $\xi = O(1)$.
If $\xi \in \ns{X}$ has the expression~\eqref{eq:xi}, then $\xi = O(\alpha)$ if and only if  $\alpha_j = O(\alpha)$ for each $j=1,\dots,n$.
For vector spaces $X, Y \in V(S)$, let $\cL (X\to Y)$ denote the set of linear maps from $X$ to $Y$.
If vector spaces $X,Y, Z \in V(S)$ are finite-dimensional and $A \in \ns{\cL(X\to Y)}$ and $B \in \ns{\cL (Y\to Z)}$ satisfy $A = O(\alpha)$ and $B = O(\beta)$ for some $\alpha, \beta \in \ns{\realn}$, then $BA = O(\alpha\beta)$.
In specific, from the case of $\alpha=\beta=1$, we conclude that if $A$ and $B$ are finite, then so is $BA$. Moreover for finite $A, A^\prime \in \ns{\cL(X\to Y)}$ and finite $B \in \ns{\cL (Y\to Z)}$ we have
\begin{equation}
	\st(A+A^\prime)=\st(A)+\st(A^\prime),\quad \st(BA) = \st(B)\st(A),
	\label{eq:st1} 
\end{equation}
and if $X$ and $Y$ are Hilbert spaces
\begin{equation}
	\st(A^\dag) = \st(A)^\dag .
	\notag 
\end{equation}
Similarly for $A \in \ns{\LX}$ and $\xi \in \ns{X}$ with $A = O(\alpha)$ and $\xi = O(\beta)$, we have $A\xi = O(\alpha\beta)$ and similar equations as \eqref{eq:st1} hold.

The notions of the continuity and the differentiability of a scalar or vector-valued functions can be restated in the following nonstandard manners.
Let $X\in V(S)$ be a finite-dimensional vector space over $\cmplx$, let $I \subseteq \realn$ be an interval which may be closed or open in the right and left, and let $F \colon I \to X$ be a vector-valued function.
Then $F$ is continuous at $t \in I$ if and only if for any $s \in \ns{I}$, $s \approx t$ implies $\ns{F}(s) \approx F(t)$, that is $\ns{F}$ maps any point infinitely close to $t$ to a point infinitely close to $F(t)$.
$F$ is differentiable at $t \in I$ and $F^\prime(t) = c \in \realn$ if and only if $c = \st\left(\frac{\ns{F}(s) -F(t)}{s-t}\right)$ for any $s \in \ns{I}$ with $s\approx t$ and $s\neq t$.

We also have a nonstandard characterization of the limit of a sequence.
Let $(a_n)_{n\in \natn}$ be a sequence in a finite-dimensional vector space $X \in V(S)$.
Then $a_n$ converges to $a \in X$ if and only if $\ns{a}_\nu \approx a$ for any infinite hypernatural number $\nu \in \ns{\natn}$.

\subsection{CP map and Kraus representation}
Let $\cH$ be a finite-dimensional complex Hilbert space.
A linear map $\Psi \in \LH$ is called \textit{positive} if $\Psi(A) \geq O$ for any $A \in \LH$ with $A\geq O$, where $O\in \LH$ is the zero operator and for self-adjoint $B,C\in \LH$ we defined the matrix order $B\geq C$ $:\defarrow$ $\braket{\psi|(B-C)\psi} \geq 0$ $(\forall \psi \in \cH)$.
$\Psi$ is called CP if for each $n \in \natn$ the map $\Psi \otimes \cI_n \in \cL (\cL (\cH \otimes \cmplx^n))$ is positive, where $\cI_n$ denotes the identity map on $\mathcal{L}(\cmplx^n)$.
It is well-known~\cite{%
Kraus1971311,davies1976quantum,kraus10.1007/3-540-12732-1,nielsenchuang%
} that $\Psi \in \LLH$ is CP if and only if $\Psi$ has the following \textit{Kraus representation}
\[
	\Psi(A) = \sum_{j=1}^N M_j^\dag A M_j, \quad (A\in \LH)
\]
for some $M_1,\dots ,M_N \in \LH$.
The operators $(M_j)_{j=1}^N$ are called \textit{Kraus operators} of $\Psi$.
The number $N$ of Kraus operators can be chosen to be at most $d^2$, where $d = \dim \cH$.
If the map $\Psi$ further satisfies the unitality $\Psi(\unit)=\unit$, then the Kraus operators satisfy the completeness condition\[
	\sum_{j=1}^N M_j^\dag M_j =\unit.
\]
In this paper we only use the existence of Kraus representation instead of the original definition of the complete positivity.

\section{Proof of Theorem~\ref{thm:gksl}} \label{sec:proof}
We first assume that $\Lambda_t = \exp(tL)$ $(t\geq 0)$ is a QDS and show that $L$ is a GKSL generator with $d^2$ traceless jump operators and traceless $H$.
From the assumption $\Lambda_t$ has Kraus operators satisfying the completeness condition for each real number $t \geq 0$.
By the transfer principle, for each $0 \leq \tau \in  \ns{\realn}$ there exist $\ast$-operators $L_0(\tau) , L_1(\tau), \dots, L_{d^2-1} (\tau) \in \ns{\LH}$ such that
\[
	\Lambda_\tau (A) = \sum_{j=0}^{d^2-1} L_j (\tau)^\dag A L_j(\tau) \quad (A \in \ns{\LH})
\]
and 
\begin{equation}
	\Lambda_{\tau}(\unit) = \sum_{j=0}^{d^2-1} L_j(\tau)^\dag L_j(\tau) = \unit .
	\label{eq:unitalK}
\end{equation}
We fix an infinitesimal hyperreal number $\dt >0$ and define traceless $\ast$-operators
\[
	M_j := L_j(\dt) - \alpha_j \unit , \quad \alpha_j:= d^{-1} \tr[L_j(\dt)] .
\]
Then from the completeness condition~\eqref{eq:unitalK} we have
\begin{align*}
	\unit &= \sum_{j=0}^{d^2-1} (M_j^\dag + \overline{\alpha_j}\unit )(M_j + \alpha_j \unit) 
	\\
	&= \sum_{j=0}^{d^2-1} M_j^\dag M_j + X+X^\dag + \beta \unit ,
\end{align*}
where 
\[
	X := \sum_{j=0}^{d^2-1} \overline{\alpha_j} L_j, \quad \beta := \sum_{j=0}^{d^2-1} |\alpha_j|^2 .
\]
Hence we obtain
\[
	X_R := \frac{X+X^\dag}{2} = \frac{1}{2} (1-\beta)\unit  - \frac{1}{2} \sum_{j=1}^{d^2-1} M_j^\dag M_j .
\]
Thus for every $A\in \ns{\LH}$ we have
\begin{align}
	\Lambda_{\dt} (A) 
	&= \sum_{j=0}^{d^2-1} (M_j^\dag + \overline{\alpha_j}\unit )A(M_j + \alpha_j \unit) \notag \\
	&= X^\dag A + AX + \beta A +  \sum_{j=0}^{d^2-1} M_j^\dag A M_j \notag \\
	&= -i[X_I,A] + \{X_R,A \} + \beta A +  \sum_{j=0}^{d^2-1} M_j^\dag A M_j \quad \left( X_I := \frac{X-X^\dag}{2i}\right) \notag \\
	&= -i[X_I,A] + \frac{1}{2} \left\{ (1-\beta)\unit  - \sum_{j=1}^{d^2-1} M_j^\dag M_j,A \right\} + \beta A +  \sum_{j=0}^{d^2-1} M_j^\dag A M_j
	\notag \\
	&= A -i[X_I,A] +\sum_{j=0}^{d^2-1}\left(M_j^\dag A M_j - \frac{1}{2} \{M_j^\dag M_j , A \}  \right)  \notag \\
	&= A + i[Y,A] + \sum_{j=0}^{d^2-1} \left( M_j^\dag A M_j - \frac{1}{2}  \{ M_j^\dag M_j, A  \} \right) ,
	\label{eq:4}
\end{align}
where we defined
\[
	Y := -X_I + d^{-1} \tr[X_I] \unit ,
\]
which is traceless and self-adjoint.

Now from 
\begin{equation}
	\Gamma_{\dt} := \Lambda_{\dt} - \cI = O(\dt ),
	\notag
\end{equation}
for any finite $A \in \ns{\LH}$ we have
\begin{equation}
	\Gamma_{\dt} (A) = i[Y,A] + \sum_{j =0}^{d^2-1} \left( M_j^\dag A M_j - \frac{1}{2} \{ M_j^\dag M_j, A \} \right) = O(\dt).
	\label{eq:Gamma}
\end{equation}
Then from \eqref{eq:Gamma}, for an orthonormal basis $(e_k)_{k=1}^d$ of $\cH$ 
\begin{align}
	O(\dt) &=
	\sum_{k,\ell =1}^d \braket{e_k| \Gamma_{\dt}(\ket{e_k}\bra{e_\ell})e_\ell}
	\label{eq:trGdt} 
	\\
	&=\cancel{ i \sum_{k,\ell =1}^d \braket{e_k |Ye_k } \braket{e_\ell |e_\ell}} -\cancel{ i \sum_{k,\ell =1}^d \braket{e_k|e_k} \braket{e_\ell|Ye_\ell} } \notag \\
	&\quad + \sum_{j=0}^{d^2-1} \sum_{k,\ell=1}^d \left( \braket{e_k|M_j^\dag e_k} \braket{e_\ell |M_j e_\ell} -\frac{1}{2} \braket{e_k|M_j^\dag M_j e_k} \braket{e_\ell|e_\ell} - \frac{1}{2} \braket{e_k|e_k} \braket{e_\ell|M_j^\dag M_j e_\ell} \right) \notag  \\
	&= \sum_{j=0}^{d^2-1} \left( \tr[M_j^\dag] \tr[M_j] - d \tr[M_j^\dag M_j]  \right) \notag  \\
	&= -d \sum_{j=0}^{d^2-1}  \| M_j \|_2^2, \notag 
\end{align}
where $\| A \|_2 := \sqrt{\tr[A^\dag A]}$ is the Hilbert-Schmidt norm and we used $\tr[M_j]=0$ in the last equality.
From this we have
\[
	0 \leq \| M_j \|_2 \leq \sqrt{\sum_{k=0}^{d^2-1} \| M_k \|_2^2} = O(\sqrt{\dt})
\]
and hence
\begin{equation}
	M_j = O(\sqrt{\dt}).
	\label{eq:NjO}
\end{equation}

We next estimate the order of $Y$.
Let $y_\rmax$ and $y_\rmin$ be the maximum and minimum $\ast$-eigenvalues of $Y$, respectively, and
let $\psi_{\mathrm{max}}, \psi_{\mathrm{min}} \in \ns{\cH}$ be normalized $\ast$-eigenvectors of $Y$ with $Y \psi_{\rmax ( \rmin )} = y_{\rmax( \rmin )} \psi_{\rmax ( \rmin )}$.
From $\tr [Y] =0$, we have 
\[
  y_\rmin \leq 0 \leq y_\rmax, \quad \|Y\|_\infty = \max(y_\rmax, - y_\rmin) ,
\]
where 
\[
\| A\|_\infty := \sup_{\psi \in \cH \colon \| \psi\|=1} \| A\psi\|
\]
is the uniform norm with respect to the norm $\| \psi \| = \braket{\psi|\psi}^{1/2}$ on $\cH$.
From \eqref{eq:Gamma} and \eqref{eq:NjO} we have 
\[
	[Y,A] = O(\dt)
\]
for any finite $A \in \ns{\LH}$ and hence
\begin{align*}
	O(\dt) &= \| [Y, \ket{\psi_\rmax}\bra{\psi_\rmin}]\|_\infty \\
	&=|y_\rmax - y_\rmin| \| \ket{\psi_\rmax}\bra{\psi_\rmin} \|_\infty \\
	&= y_\rmax -y_\rmin \\
	& \geq \max(y_\rmax, - y_\rmin) \\
	&=\|Y\|_\infty .
\end{align*}
Thus we obtain
\begin{equation}
	Y = O(\dt) .
	\label{eq:ZO}
\end{equation}

From \eqref{eq:NjO} and \eqref{eq:ZO}, standard operators
\begin{equation}
		H := \st\left( \dt^{-1} Y \right), \quad W_j :=  \st\left( \dt^{-1/2} M_j \right)
		\label{eq:jdef}
\end{equation}
are well-defined.
Then the operators $H=H^\dag$ and $W_j$ are traceless.
Moreover, for each $A\in \LH$, $L(A)$ is given by
\begin{align}
	L(A) &= \left. \frac{d}{dt} \Lambda_t(A) \right|_{t=0} \notag \\
	&=\st\left( \frac{\Lambda_{\dt}(A) - A}{\dt} \right) \notag \\
	&= \st\left(
	 i[\dt^{-1}Y,A] + \sum_{j =0}^{d^2-1} \left[ (\dt^{-1/2}M_j)^\dag A (\dt^{-1/2}M_j) - \frac{1}{2} \left\{ (\dt^{-1/2}M_j)^\dag (\dt^{-1/2}M_j), A \right\} \right]
	\right) \notag \\
	&= i[H,A] + \sum_{j =0}^{d^2-1} \left( W_j^\dag A W_j - \frac{1}{2}  \{ W_j^\dag W_j, A \} \right).\label{eq:LA}
\end{align}
Thus $L$ is a GKSL generator with $d^2$ traceless jump operators.\footnote{
The rest of the proof is almost the same as the standard one except the introduction of an infinite hypernatural number $\nu$ in the converse part, while we include this for completeness.
}

We next show that the number of jump operators can be reduced to $d^2-1$.
Since 
\[
	\dim \{A \in {\LH}| \tr[A] = 0\} = d^2-1,
\]
the traceless operators $(W_j)_{j=0}^{d^2-1}$ are linearly dependent.
Hence there exist complex numbers $(c_j)_{j=0}^{d^2-1}$ such that
\[
	\sum_{j=0}^{d^2-1} c_j W_j = O 
\]
and $c_j \neq 0$ for some $j$.
By multiplying a nonzero constant to $c_j$\rq{}s, we may assume without loss of generality that \[ \sum_{j=0}^{d^2-1}|c_j|^2 =1. \]
We take a unitary matrix $(u_{kj})_{k,j=0}^{d^2-1}$ with complex entries such that 
\[u_{0j} = c_j \quad (j=0,1,\dots, d^2-1)\] and define traceless operators
\[
	V_k := \sum_{j=0}^{d^2-1} u_{kj}W_j.
\]
Then from 
\[
	W_j = \sum_{k=0}^{d^2-1} \overline{u_{kj}} V_k
\]
and the expression \eqref{eq:LA}, we obtain
\begin{align}
	L(A) &= 
	i[H,A] + \sum_{j,k,\ell =0}^{d^2-1} u_{kj} \overline{u_{\ell j}} \left( V_k^\dag A V_\ell - \frac{1}{2} \{ V_k^\dag V_\ell, A \} \right) \notag
	\\
	&= i[H,A] + \sum_{k,\ell =0}^{d^2-1} \delta_{k \ell} \left( V_k^\dag A V_\ell - \frac{1}{2}  \{ V_k^\dag V_\ell, A  \} \right) \notag \\
	&= i[H,A] + \sum_{j =1}^{d^2-1} \left( V_j^\dag A V_j - \frac{1}{2} \{ V_j^\dag V_j, A  \} \right) ,\notag
\end{align}
where in the last equality we used
\[
	V_0 = \sum_{j=0}^{d^2-1} c_j W_j = O.
\]
Thus the first paragraph of Theorem~\ref{thm:gksl} is proved.

We now assume that $L$ is a GKSL generator given by \eqref{eq:gksl1} and show that $\Lambda_t = \exp(tL)$ $(t\geq 0)$ is a QDS.
We take a sufficiently small real number $t_0 >0$ such that
\[
	t_0 \sum_{j=1}^N V_j^\dag V_j  \leq \unit
\]
(e.g.\ $t_0 = \| \sum_{j=1}^N V_j^\dag V_j \|_\infty^{-1}$ if some $V_j$ is nonzero) and for each $t \in [0,t_0]$ we define a unital CP map $\Psi_t \in \LLH $ by
\begin{align*}
	\Psi_t (A)&:= \sum_{j=0}^N R_j(t)^\dag AR_j(t) \quad (A\in \LH),\\
	R_j(t) &:= 
	\begin{cases}
	e^{-itH} \sqrt{\unit -t \sum_{j=1}^N V_j^\dag V_j} &(j=0), \\
	\sqrt{t} V_j &(j=1,\dots, N).
	\end{cases}
\end{align*}
Take an arbitrary real number $t\geq 0$ and define unital CP map $\Phi_n := \Psi_{t/n}^{n}$ for each $n\in \natn$ with $n\geq t/t_0$. 
Then, by the transfer principle, for any $n \in \ns{\natn}$ with $n\geq t/t_0$ the $\ast$-map $\Phi_n$ is unital and has Kraus $\ast$-operators satisfying the completeness condition.
We fix an infinite hypernatural number $\nu\in \ns{\natn}$.
From 
\begin{align*}
	R_0(t/\nu) &= e^{-i\frac{t}{\nu}H} \sqrt{\unit -\frac{t}{\nu} \sum_{j=1}^N V_j^\dag V_j} \\
	&= \unit - \frac{ it}{\nu} H - \frac{t}{2\nu} \sum_{j=1}^N V_j^\dag V_j + O(\nu^{-2}),
\end{align*}
for any finite $A\in \ns{\LH}$ we have
\begin{align*}
	\Psi_{t/\nu} (A) &= \left(\unit - \frac{ it}{\nu} H - \frac{t}{2\nu} \sum_{j=1}^N V_j^\dag V_j  \right)^\dag A \left(\unit - \frac{ it}{\nu} H - \frac{t}{2\nu} \sum_{j=1}^N V_j^\dag V_j \right) + \frac{t}{\nu} \sum_{j=1}^N V_j^\dag A V_j + O(\nu^{-2}) \\
	&= A + \frac{t}{\nu} L(A) + O(\nu^{-2})
\end{align*}
and hence
\begin{align}
	\Phi_{\nu} &= \left(\cI + \frac{t}{\nu} L + O(\nu^{-2})  \right)^\nu \notag \\
	& \approx \exp (tL) \notag \\
	&= \Lambda_t. \notag
\end{align}
Thus we have $\Lambda_t = \st (\Phi_\nu)$.
Moreover, $\Phi_\nu$ has the Kraus representation 
\[
	\Phi_\nu (A) = \sum_{j=0}^{d^2-1} S_j^\dag A S_j, \quad S_j \in \ns{\LH}
\]
with the completeness condition
\begin{equation}
	\sum_{j=0}^{d^2-1} S_j^\dag S_j = \unit ,
	\notag
\end{equation}
from which it follows that each $S_j$ is finite.
Thus for each $A \in \LH$ we obtain
\begin{align*}
	\Lambda_t(A) &= \st\left( \sum_{j=0}^{d^2-1} S_j^\dag A S_j \right) \\
	&= \sum_{j=0}^{d^2-1} \st(S_j)^\dag A \st(S_j)
\end{align*}
and therefore $\Lambda_t$ is CP.
Moreover 
\begin{align*}
	\Lambda_t(\unit) &= \st (\Phi_\nu (\unit)) \\ &= \st(\unit) \\ &= \unit
\end{align*}
and hence $\Lambda_t$ is unital.
Thus $(\Lambda_t)_{t\geq  0}$ is a QDS. 
\qed

\section{Remarks} \label{sec:rem}
\begin{enumerate}
\item
The crucial step of our proof is the order estimation of $\ast$-operators $M_j$ and $Y$, especially the evaluation of the quantity~\eqref{eq:trGdt},
which is in fact the trace $\tr[\Gamma_{\dt}]$.
This can be shown as follows:
Since the family of operators $E_{k\ell} := \ket{e_k}\bra{e_\ell} \in \LH$ $(k,\ell =1,\dots ,d)$ is an orthonormal basis of $\LH$ equipped with the Hilbert-Schmidt inner product $\braket{A,B}_2 := \tr[A^\dag B]$, the trace of any linear map $\Psi \in \LLH$ is given by
\begin{align*}
	\tr[\Psi] &= \sum_{k,\ell=1}^d \braket{E_{k\ell}, \Psi(E_{k\ell})}_2 \\
	&= \sum_{k,\ell=1}^d \braket{e_k| \Psi(\ket{e_k}\bra{e_\ell})e_\ell}.
\end{align*}
The corresponding finite quantity $\tr[L] = -d\sum_{j}\tr[V_j^\dag V_j]$ (for traceless jump operators $V_j$) appears in Ref.~\onlinecite{kimura2017} and can be written as the minus of the summation of the relaxation rates of the QDS.
(The author would like to thank Gen Kimura for pointing out this fact.)
\item
As mentioned in the introduction, it is not obvious whether a CP map-valued function $t\mapsto \Psi_t \in \LH$ has Kraus operators $M_j(t)$ $j=1,\dots,d^2$ that are continuous in $t$.
In fact, we can show a weaker statement in a general setup that roughly claims that close CP maps have close Kraus operators. 
See Appendix~\ref{sec:app} for the statement (Proposition~\ref{prop:conti}) and its nonstandard proof.
\item
In the theory of measurement in continuous time~\cite{barchielli2009quantum,CBO9780511813948}, heuristic arguments on the measurement operators in an infinitesimal time often appear.
If we notice that there are some applications of nonstandard methods to classical stochastic analysis~\cite{stroyan1986,nelson1987,sergio2009,loebwolff2015}, nonstandard approaches may have potential applications to such continuous measurement processes.
\end{enumerate}

\begin{acknowledgments}
The author would like to thank Hayato Arai, Gen Kimura, Jaeha Lee, Shintaro Minagawa, Kanta Sonoda, Ryo Takakura, Akane Watanabe, and Yuriko Yamamoto for helpful discussions and comments.
This work was supported by JSPS Grant-in-Aid for Early-Career Scientists No.~JP22K13977.
\end{acknowledgments}

\appendix
\section{Continuity of Kraus operators} \label{sec:app}
In this appendix we prove Proposition~\ref{prop:conti} shown below.
For a finite-dimensional complex Hilbert space $\cH$ we denote by $\cph$ the set of CP maps on $\LH$ which is equipped with the relative topology of the norm topology on $\LLH$.

\begin{prop} \label{prop:conti}
Let $\cH$ be a finite-dimensional complex Hilbert space and let $\| \cdot \|$ be any norm on $\LH$.
Suppose that a CP map $\Psi \in \cph$ has the Kraus representation 
\begin{equation}
	\Psi  (A) = \sum_{j=1}^N M_j^\dag A M_j \quad (A\in \LH)
	\label{eq:PsiKraus}
\end{equation}
with $N \geq d^2$, where $d= \dim \cH$.
Then for any real number $\epsilon >0$ there exists an open neighborhood $U \subseteq \cph$ of $\Psi$ such that every $\Phi \in U$ has Kraus operators $(N_j)_{j=1}^N$ such that $\|M_j -N_j\| < \epsilon$ for $j=1,\dots ,N$.
\end{prop}
Note that we do not lose the generality by requiring $N \geq d^2$ because when $N<d^2$ we may define $M_j = O$ for $j=N+1,\dots, d^2$.

For the proof we use the following notion of infinitesimal neighborhood.
Let $X \in V(S)$ be a topological space and let $\mathscr	{T} \subseteq \mathscr{P}(X)$ be the topology on $X$, i.e.\ the family of open sets.
We denote by $\mathscr{T}(a)$ the family of open neighborhoods of $a \in X$.
From the construction of $V(S)$, $\mathscr{T}$, any $\mathscr{T}(a)$ for $a\in X$, and any element of $\mathscr{T}$ are in $V(S)$.
For $a\in X$ we define the \textit{monad} 
\[
	\mu(a) := \bigcap_{U \in \mathscr{T}(a)} \ns{U} \subseteq \ns{X}
\]
and any element $\alpha \in \mu(a)$ is said to be infinitely close to $a$ and if so we write as $\alpha \approx a$.
By the concurrence argument we can show that there exists a $\ast$-open neighborhood $T \in \ns{\mathscr{T}(a)}$ satisfying $T \subseteq \mu (a)$.
Such $T$ is called an \textit{infinitesimal neighborhood} of $a$.
If $X$ is a finite-dimensional vector space equipped with the norm topology, then $\mu(a)$ for $a\in X$ is the set of $\alpha \in \ns{X}$ such that $\| \alpha -a\|$ is infinitesimal.
Thus the definition of the symbol $\approx$ is consistent with that in Section~\ref{sec:prel}.


We also use the following fact about the freedom of the Kraus operators.
\begin{lemm}[Ref.~\onlinecite{nielsenchuang}, Theorem~8.2] \label{lemm:kraus}
Let $\Phi \in \LLH$ be a CP map with Kraus representations
\begin{align*}
	\Phi(A) &= \sum_{j=1}^N B_j^\dag A  B_j \\
	&= \sum_{j=1}^N C_j^\dag A  C_j \quad (A\in \LH).
\end{align*}
Then there exists a unitary matrix $(u_{jk})_{j,k=1}^N$ with complex entries such that 
\begin{equation}
	C_j = \sum_{k=1}^N u_{jk}B_k \quad (j=1,\dots, N).
	\label{eq:CjBk}
\end{equation}
Conversely, if $(B_j)_{j=1}^N$ are Kraus operators of $\Phi$ and the operators $(C_j)_{j=1}^N$ are given by \eqref{eq:CjBk} for some unitary matrix $(u_{jk})_{j,k=1}^N$, then $(C_j)_{j=1}^N$ are also Kraus operators of $\Phi$. 
\end{lemm}

\noindent
\textit{Proof of Proposition~\ref{prop:conti}.}
By $N\geq d^2$ and the axiom of choice, there exist maps  
\[
	F_j \colon \cph \to \LH \quad (j=1,\dots ,N)
\]
such that each $\Phi \in \cph $ has the Kraus representation 
\[
	\Phi (A) = \sum_{j=1}^N F_j(\Phi)^\dag A F_j(\Phi) \quad (A\in {\LH}).
\]
Thus by the transfer principle we have
\[
	\Xi (A) = \sum_{j=1}^N (\ns{F}_j(\Xi))^\dag A (\ns{F}_j(\Xi)) \quad (A\in \ns{\LH})
\]
for each $\Xi \in \ns{\cph}$.
Take arbitrary $\Xi \in \mu(\Psi)$, where $\mu(\Psi)\subseteq \ns{\cph}$ denotes the monad.
Then $\Xi \approx \Psi$ and hence for each $A\in \LH$ we have
\begin{align*}
	\Psi (A) &= \st (\Xi(A)) \\
	&= \st\left( \sum_{j=1}^N (\ns{F}_j(\Xi))^\dag A (\ns{F}_j(\Xi))  \right).
\end{align*}
By substituting $A=\unit$ we obtain
\[
	 \st\left(\sum_{j=1}^N (\ns{F}_j(\Xi))^\dag (\ns{F}_j(\Xi)) \right) = \Psi (\unit) \in \LH,
\]
which implies that the $\ast$-operator $\sum_{j=1}^N (\ns{F}_j(\Xi))^\dag (\ns{F}_j(\Xi)) $ is finite.
Thus the finiteness of each $\ns{F}_j(\Xi)$ follows by
\begin{align*}
	\|\ns{F}_j(\Xi)\|_2^2 &= \tr[ (\ns{F}_j(\Xi))^\dag \ns{F}_j(\Xi)] \\
	&\leq \tr\left[  \sum_{k=1}^N (\ns{F}_k(\Xi))^\dag \ns{F}_k(\Xi) \right] \\
	&=O(1).
\end{align*}
Thus we have
\[
	\Psi (A) = \sum_{j=1}^N \st(\ns{F}_j(\Xi))^\dag A \st(\ns{F}_j(\Xi)) \quad (A\in \LH),
\]
i.e.\ the standard operators $\st(\ns{F}_j(\Xi))$ $(j=1,\dots,N)$ are Kraus operators of $\Psi$.
Since $\Psi $ has another Kraus representation \eqref{eq:PsiKraus}, Lemma~\ref{lemm:kraus} implies that there exists a unitary matrix $(t_{jk})_{j,k=1}^N$ with complex entries such that 
\begin{align}
	M_j &= \sum_{k=1}^N t_{jk}\st (\ns{F}_k(\Xi))
	\notag \\
	&= \st(\tM_j(\Xi)) ,\notag
\end{align}
where we defined finite $\ast$-operators
\[
	\tM_j(\Xi) := \sum_{k=1}^N t_{jk} \ns{F}_k(\Xi) \quad (j=1,\dots ,N).
\]
Then from Lemma~\ref{lemm:kraus} and the transfer principle, the $\ast$-operators $(\tM_j(\Xi))_{j=1}^N$ are Kraus $\ast$-operators of $\Xi$.
Thus we have shown that for any $\Xi \in \mu(\Psi)$ there exist Kraus $\ast$-operators $(\tM_j(\Xi))_{j=1}^N$ of $\Xi$ such that $\tM_j(\Xi) \approx M_j$.

From this, for each real number $\epsilon >0$ the following statement holds: 
There exists $U_0 \in \ns{\mathscr{T}(\Psi)}$ such that for every $\Xi \in U_0$ there exist Kraus $\ast$-operators $(A_j)_{j=1}^N$ of $\Xi$ such that $\|  A_j - M_j\| < \epsilon$ for $j=1,\dots ,N$, where $\mathscr{T}(\Psi) \subseteq \mathscr{P}(\cph)$ denotes the family of open neighborhoods of $\Psi$.
We can immediately verify this by taking an infinitesimal neighborhood $U_0 \in \ns{\mathscr{T}(\Psi)}$ with $U_0 \subseteq \mu(\Psi)$ and $A_j = \tM_j(\Xi)$.
Thus by the transfer principle, there exists $U \in \mathscr{T}(\Psi)$ such that each $\Phi \in U$ has Kraus operators $N_j \in \LH$ $(j=1,\dots ,N)$ satisfying
\begin{equation}
	\| M_j  - N_j\| < \epsilon \quad (j=1,\dots ,N), \notag
\end{equation}
which proves the claim.
\qed


\begin{thebibliography}{21}%
\makeatletter
\providecommand \@ifxundefined [1]{%
 \@ifx{#1\undefined}
}%
\providecommand \@ifnum [1]{%
 \ifnum #1\expandafter \@firstoftwo
 \else \expandafter \@secondoftwo
 \fi
}%
\providecommand \@ifx [1]{%
 \ifx #1\expandafter \@firstoftwo
 \else \expandafter \@secondoftwo
 \fi
}%
\providecommand \natexlab [1]{#1}%
\providecommand \enquote  [1]{``#1''}%
\providecommand \bibnamefont  [1]{#1}%
\providecommand \bibfnamefont [1]{#1}%
\providecommand \citenamefont [1]{#1}%
\providecommand \href@noop [0]{\@secondoftwo}%
\providecommand \href [0]{\begingroup \@sanitize@url \@href}%
\providecommand \@href[1]{\@@startlink{#1}\@@href}%
\providecommand \@@href[1]{\endgroup#1\@@endlink}%
\providecommand \@sanitize@url [0]{\catcode `\\12\catcode `\$12\catcode `\&12\catcode `\#12\catcode `\^12\catcode `\_12\catcode `\%12\relax}%
\providecommand \@@startlink[1]{}%
\providecommand \@@endlink[0]{}%
\providecommand \url  [0]{\begingroup\@sanitize@url \@url }%
\providecommand \@url [1]{\endgroup\@href {#1}{\urlprefix }}%
\providecommand \urlprefix  [0]{URL }%
\providecommand \Eprint [0]{\href }%
\providecommand \doibase [0]{http://dx.doi.org/}%
\providecommand \selectlanguage [0]{\@gobble}%
\providecommand \bibinfo  [0]{\@secondoftwo}%
\providecommand \bibfield  [0]{\@secondoftwo}%
\providecommand \translation [1]{[#1]}%
\providecommand \BibitemOpen [0]{}%
\providecommand \bibitemStop [0]{}%
\providecommand \bibitemNoStop [0]{.\EOS\space}%
\providecommand \EOS [0]{\spacefactor3000\relax}%
\providecommand \BibitemShut  [1]{\csname bibitem#1\endcsname}%
\let\auto@bib@innerbib\@empty
\bibitem [{\citenamefont {Davies}(1980)}]{davies1980}%
  \BibitemOpen
  \bibfield  {author} {\bibinfo {author} {\bibfnamefont {E.~B.}\ \bibnamefont {Davies}},\ }\href@noop {} {\emph {\bibinfo {title} {One-parameter semigroups}}}\ (\bibinfo  {publisher} {Academic Press},\ \bibinfo {year} {1980})\BibitemShut {NoStop}%
\bibitem [{\citenamefont {Gorini}, \citenamefont {Kossakowski},\ and\ \citenamefont {Sudarshan}(1976)}]{Gorini1976}%
  \BibitemOpen
  \bibfield  {author} {\bibinfo {author} {\bibfnamefont {V.}~\bibnamefont {Gorini}}, \bibinfo {author} {\bibfnamefont {A.}~\bibnamefont {Kossakowski}}, \ and\ \bibinfo {author} {\bibfnamefont {E.~C.~G.}\ \bibnamefont {Sudarshan}},\ }\bibfield  {title} {\enquote {\bibinfo {title} {{Completely positive dynamical semigroups of $N$-level systems}},}\ }\href {\doibase 10.1063/1.522979} {\bibfield  {journal} {\bibinfo  {journal} {J. Math. Phys.}\ }\textbf {\bibinfo {volume} {17}},\ \bibinfo {pages} {821--825} (\bibinfo {year} {1976})}\BibitemShut {NoStop}%
\bibitem [{\citenamefont {Lindblad}(1976)}]{lindblad1976}%
  \BibitemOpen
  \bibfield  {author} {\bibinfo {author} {\bibfnamefont {G.}~\bibnamefont {Lindblad}},\ }\bibfield  {title} {\enquote {\bibinfo {title} {On the generators of quantum dynamical semigroups},}\ }\href {\doibase 10.1007/BF01608499} {\bibfield  {journal} {\bibinfo  {journal} {Comm. Math. Phys.}\ }\textbf {\bibinfo {volume} {48}},\ \bibinfo {pages} {119--130} (\bibinfo {year} {1976})}\BibitemShut {NoStop}%
\bibitem [{\citenamefont {Davies}(1976)}]{davies1976quantum}%
  \BibitemOpen
  \bibfield  {author} {\bibinfo {author} {\bibfnamefont {E.~B.}\ \bibnamefont {Davies}},\ }\href@noop {} {\emph {\bibinfo {title} {Quantum theory of open systems}}}\ (\bibinfo  {publisher} {Academic Press},\ \bibinfo {year} {1976})\BibitemShut {NoStop}%
\bibitem [{\citenamefont {Stinespring}(1955)}]{1955stinespring}%
  \BibitemOpen
  \bibfield  {author} {\bibinfo {author} {\bibfnamefont {W.~F.}\ \bibnamefont {Stinespring}},\ }\bibfield  {title} {\enquote {\bibinfo {title} {{Positive functions on $C^\ast$-algebras}},}\ }\href {http://www.jstor.org/stable/2032342} {\bibfield  {journal} {\bibinfo  {journal} {Proc. Amer. Math. Soc.}\ }\textbf {\bibinfo {volume} {6}},\ \bibinfo {pages} {211--216} (\bibinfo {year} {1955})}\BibitemShut {NoStop}%
\bibitem [{\citenamefont {Kraus}(1971)}]{Kraus1971311}%
  \BibitemOpen
  \bibfield  {author} {\bibinfo {author} {\bibfnamefont {K.}~\bibnamefont {Kraus}},\ }\bibfield  {title} {\enquote {\bibinfo {title} {General state changes in quantum theory},}\ }\href {\doibase 10.1016/0003-4916(71)90108-4} {\bibfield  {journal} {\bibinfo  {journal} {Ann. Phys. (NY)}\ }\textbf {\bibinfo {volume} {64}},\ \bibinfo {pages} {311 -- 335} (\bibinfo {year} {1971})}\BibitemShut {NoStop}%
\bibitem [{\citenamefont {Kraus}(1983)}]{kraus10.1007/3-540-12732-1}%
  \BibitemOpen
  \bibfield  {author} {\bibinfo {author} {\bibfnamefont {K.}~\bibnamefont {Kraus}},\ }\href {\doibase 10.1007/3-540-12732-1} {\emph {\bibinfo {title} {States, Effects, and Operations}}}\ (\bibinfo  {publisher} {Berlin: Springer},\ \bibinfo {year} {1983})\BibitemShut {NoStop}%
\bibitem [{\citenamefont {Chru\'{s}ci\'{n}ski}\ and\ \citenamefont {Pascazio}(2017)}]{chruscinski2017}%
  \BibitemOpen
  \bibfield  {author} {\bibinfo {author} {\bibfnamefont {D.}~\bibnamefont {Chru\'{s}ci\'{n}ski}}\ and\ \bibinfo {author} {\bibfnamefont {S.}~\bibnamefont {Pascazio}},\ }\bibfield  {title} {\enquote {\bibinfo {title} {A brief history of the {GKLS} equation},}\ }\href {\doibase 10.1142/S1230161217400017} {\bibfield  {journal} {\bibinfo  {journal} {Open Systems \& Information Dynamics}\ }\textbf {\bibinfo {volume} {24}},\ \bibinfo {pages} {1740001} (\bibinfo {year} {2017})}\BibitemShut {NoStop}%
\bibitem [{\citenamefont {Kossakowski}(1972)}]{kossakowski1972necessary}%
  \BibitemOpen
  \bibfield  {author} {\bibinfo {author} {\bibfnamefont {A.}~\bibnamefont {Kossakowski}},\ }\bibfield  {title} {\enquote {\bibinfo {title} {On necessary and sufficient conditions for a generator of a quantum dynamical semi-group},}\ }\href@noop {} {\bibfield  {journal} {\bibinfo  {journal} {Bull. Acad. Pol. Sci. S\'er. Math. Astr. Phys.}\ } (\bibinfo {year} {1972})}\BibitemShut {NoStop}%
\bibitem [{\citenamefont {Preskill}(1998)}]{preskill1998lecture}%
  \BibitemOpen
  \bibfield  {author} {\bibinfo {author} {\bibfnamefont {J.}~\bibnamefont {Preskill}},\ }\href {http://theory.caltech.edu/~preskill/ph229/} {\enquote {\bibinfo {title} {Lecture notes for physics 229: Quantum information and computation},}\ }\bibinfo {howpublished} {California Institute of Technology, Pasadena, CA, U.S.A.} (\bibinfo {year} {1998})\BibitemShut {NoStop}%
\bibitem [{\citenamefont {Davis}(1977)}]{davis1977}%
  \BibitemOpen
  \bibfield  {author} {\bibinfo {author} {\bibfnamefont {M.}~\bibnamefont {Davis}},\ }\href {https://cir.nii.ac.jp/crid/1130000797453529856} {\emph {\bibinfo {title} {Applied nonstandard analysis}}}\ (\bibinfo  {publisher} {Wiley},\ \bibinfo {year} {1977})\BibitemShut {NoStop}%
\bibitem [{\citenamefont {Loeb}\ and\ \citenamefont {Wolff}(2015)}]{loebwolff2015}%
  \BibitemOpen
  \bibfield  {author} {\bibinfo {author} {\bibfnamefont {P.}~\bibnamefont {Loeb}}\ and\ \bibinfo {author} {\bibfnamefont {M.}~\bibnamefont {Wolff}},\ }\href {https://cir.nii.ac.jp/crid/1130282271208803584} {\emph {\bibinfo {title} {Nonstandard analysis for the working mathematician}}},\ \bibinfo {edition} {2nd}\ ed.\ (\bibinfo  {publisher} {Springer},\ \bibinfo {year} {2015})\BibitemShut {NoStop}%
\bibitem [{\citenamefont {Conway}(1990)}]{conway1990}%
  \BibitemOpen
  \bibfield  {author} {\bibinfo {author} {\bibfnamefont {J.~B.}\ \bibnamefont {Conway}},\ }\href {https://cir.nii.ac.jp/crid/1130282273040025088} {\emph {\bibinfo {title} {A course in functional analysis}}},\ \bibinfo {edition} {2nd}\ ed.,\ Graduate texts in mathematics\ (\bibinfo  {publisher} {Springer},\ \bibinfo {year} {1990})\BibitemShut {NoStop}%
\bibitem [{\citenamefont {Nielsen}\ and\ \citenamefont {Chuang}(2010)}]{nielsenchuang}%
  \BibitemOpen
  \bibfield  {author} {\bibinfo {author} {\bibfnamefont {M.~A.}\ \bibnamefont {Nielsen}}\ and\ \bibinfo {author} {\bibfnamefont {I.~L.}\ \bibnamefont {Chuang}},\ }\href@noop {} {\emph {\bibinfo {title} {Quantum computation and quantum information}}},\ \bibinfo {edition} {10th}\ ed.\ (\bibinfo  {publisher} {Cambridge University Press},\ \bibinfo {year} {2010})\BibitemShut {NoStop}%
\bibitem [{Note1()}]{Note1}%
  \BibitemOpen
  \bibinfo {note} {The rest of the proof is almost the same as the standard one except the introduction of an infinite hypernatural number $\nu $ in the converse part, while we include this for completeness.}\BibitemShut {Stop}%
\bibitem [{\citenamefont {Kimura}, \citenamefont {Ajisaka},\ and\ \citenamefont {Watanabe}(2017)}]{kimura2017}%
  \BibitemOpen
  \bibfield  {author} {\bibinfo {author} {\bibfnamefont {G.}~\bibnamefont {Kimura}}, \bibinfo {author} {\bibfnamefont {S.}~\bibnamefont {Ajisaka}}, \ and\ \bibinfo {author} {\bibfnamefont {K.}~\bibnamefont {Watanabe}},\ }\bibfield  {title} {\enquote {\bibinfo {title} {Universal constraints on relaxation times for $d$-level {GKLS} master equations},}\ }\href {\doibase 10.1142/S1230161217400091} {\bibfield  {journal} {\bibinfo  {journal} {{Open Systems \& Information Dynamics}}\ }\textbf {\bibinfo {volume} {24}},\ \bibinfo {pages} {1740009} (\bibinfo {year} {2017})}\BibitemShut {NoStop}%
\bibitem [{\citenamefont {Barchielli}\ and\ \citenamefont {Gregoratti}(2009)}]{barchielli2009quantum}%
  \BibitemOpen
  \bibfield  {author} {\bibinfo {author} {\bibfnamefont {A.}~\bibnamefont {Barchielli}}\ and\ \bibinfo {author} {\bibfnamefont {M.}~\bibnamefont {Gregoratti}},\ }\href@noop {} {\emph {\bibinfo {title} {Quantum trajectories and measurements in continuous time: the diffusive case}}},\ Lecture notes in physics\ (\bibinfo  {publisher} {Springer},\ \bibinfo {year} {2009})\BibitemShut {NoStop}%
\bibitem [{\citenamefont {Wiseman}\ and\ \citenamefont {Milburn}(2010)}]{CBO9780511813948}%
  \BibitemOpen
  \bibfield  {author} {\bibinfo {author} {\bibfnamefont {H.~M.}\ \bibnamefont {Wiseman}}\ and\ \bibinfo {author} {\bibfnamefont {G.~J.}\ \bibnamefont {Milburn}},\ }\href {\doibase 10.1017/CBO9780511813948} {\emph {\bibinfo {title} {Quantum Measurement and Control}}}\ (\bibinfo  {publisher} {Cambridge University Press},\ \bibinfo {year} {2010})\BibitemShut {NoStop}%
\bibitem [{\citenamefont {Stroyan}\ and\ \citenamefont {Bayod}(1986)}]{stroyan1986}%
  \BibitemOpen
  \bibfield  {author} {\bibinfo {author} {\bibfnamefont {K.~D.}\ \bibnamefont {Stroyan}}\ and\ \bibinfo {author} {\bibfnamefont {J.~M.}\ \bibnamefont {Bayod}},\ }\href {https://cir.nii.ac.jp/crid/1130000798344058112} {\emph {\bibinfo {title} {Foundations of infinitesimal stochastic analysis}}},\ Studies in logic and the foundations of mathematics\ (\bibinfo  {publisher} {North-Holland},\ \bibinfo {year} {1986})\BibitemShut {NoStop}%
\bibitem [{\citenamefont {Nelson}(1987)}]{nelson1987}%
  \BibitemOpen
  \bibfield  {author} {\bibinfo {author} {\bibfnamefont {E.}~\bibnamefont {Nelson}},\ }\href {https://cir.nii.ac.jp/crid/1130000798176953856} {\emph {\bibinfo {title} {Radically elementary probability theory}}},\ Annals of mathematics studies\ (\bibinfo  {publisher} {Princeton University Press},\ \bibinfo {year} {1987})\BibitemShut {NoStop}%
\bibitem [{\citenamefont {Albeverio}\ \emph {et~al.}(1986)\citenamefont {Albeverio}, \citenamefont {Fenstad}, \citenamefont {H{\o}egh-Krohn},\ and\ \citenamefont {Lindstr{\o}m}}]{sergio2009}%
  \BibitemOpen
  \bibfield  {author} {\bibinfo {author} {\bibfnamefont {S.}~\bibnamefont {Albeverio}}, \bibinfo {author} {\bibfnamefont {J.~E.}\ \bibnamefont {Fenstad}}, \bibinfo {author} {\bibfnamefont {R.}~\bibnamefont {H{\o}egh-Krohn}}, \ and\ \bibinfo {author} {\bibfnamefont {T.}~\bibnamefont {Lindstr{\o}m}},\ }\href {https://cir.nii.ac.jp/crid/1130282271090818944} {\emph {\bibinfo {title} {Nonstandard methods in stochastic analysis and mathematical physics}}}\ (\bibinfo  {publisher} {Academic Press},\ \bibinfo {year} {1986})\BibitemShut {NoStop}%
\end{thebibliography}

%

\end{document}